\documentclass[preprint,aps]{revtex4}

\usepackage{graphicx}

\begin{document}

\title{
Evidence for Multiple Underlying Fermi Surface and Isotropic Energy Gap in the Cuprate Parent Compound Ca$_2$CuO$_2$Cl$_2$}

 \author{Cheng Hu$^{1,2}$, Jian-Fa Zhao$^{1,2}$, Ying Ding$^{1,2}$, Jing Liu$^{1,2}$, Qiang Gao$^{1,2}$, Lin Zhao$^{1}$, Guo-Dong Liu$^{1}$, Li Yu$^{1}$, Chang-Qing Jin$^{1,2,4}$, Chuang-Tian Chen$^{3}$, Zu-Yan Xu$^{3}$  and Xing-Jiang Zhou$^{1,2,4,*}$}

\affiliation{
\\$^{1}$Beijing National Laboratory for Condensed Matter Physics, Institute of Physics, Chinese Academy of Sciences, Beijing 100190, China
\\$^{2}$University of Chinese Academy of Sciences, Beijing 100049, China
\\$^{3}$Technical Institute of Physics and Chemistry, Chinese Academy of Sciences, Beijing 100190, China
\\$^{4}$Collaborative Innovation Center of Quantum Matter, Beijing 100871, China
}
\date{May 28, 2018}

\maketitle

%
%

{\bf The parent compounds of the  high-temperature cuprate superconductors are Mott insulators. It has been generally agreed that understanding the physics of the doped Mott insulators is essential to understanding the mechanism of high temperature superconductivity. A natural starting point is to elucidate the basic electronic structure of the parent compound. Here we report comprehensive high resolution angle-resolved photoemission measurements on  Ca$_2$CuO$_2$Cl$_2$, a Mott insulator and a prototypical parent compound of the cuprates. Multiple underlying Fermi surface sheets are revealed for the first time. The high energy waterfall-like band dispersions exhibit different behavior near the nodal and antinodal regions. Two distinct energy scales are  identified: a \emph{d}-wave-like low energy peak dispersion and a nearly isotropic lower Hubbard band gap. These observations provide new information on the electronic structure of the cuprate parent compound, which is important for understanding the anomalous physical properties and superconductivity mechanism of the high temperature cuprate superconductors.
}


The parent compounds of the  high temperature cuprate superconductors are known as Mott insulators\cite{Anderson1991}. These materials have one unpaired electron per Cu site, which are predicted to be half-filled metal by band theory. The large onsite Coulomb repulsion, however, prohibits electron hopping between neighboring sites and leads to an insulating ground state with antiferromagnetic ordering. When charge carriers are introduced into the CuO$_2$ planes, the insulating phase is destroyed and the system turns into a metal. Upon further doping, superconductivity emerges. It has been suggested that the physics of high temperature superconductivity is  doping a Mott insulator\cite{Lee2006}. A natural starting point is to elucidate the basic electronic structure of the parent Mott insulator, i.e., to study the doped single hole behavior in the presence of strong electron correlations.


The oxychloride cuprate Ca$_2$CuO$_2$Cl$_2$ (CCOC) is a half-filled Mott insulator which has the crystal structure of La$_2$CuO$_4$\cite{Miller1990}. It can be doped into a high-temperature superconductor by replacing Ca with Na\cite{Hiori1995}. Optical spectroscopy\cite{Waku2006} and scanning tunnelling microscopy (STM)\cite{Ye2013} measurements reveal a charge transfer gap to be $\sim$2.0~eV for CCOC. Angle-resolved photoemission (ARPES) measurements identify a broad low-energy charge-transfer band feature in the spectra, which is commonly referred to as the Zhang-Rice singlet\cite{Wells1995}. It has a bandwidth of $\sim$0.3~eV and can be described well by the extended \emph{t-J} model\cite{Kim1998}. In addition, the low-energy peak exhibits a {\it d}-wave-like dispersion and forms a single remnant Fermi surface in the Mott gap region\cite{Ronning1998}. Later in this parent compound, ARPES measurements reveal a high energy kink feature near 0.8~eV along the nodal direction and a rapidly dispersing high energy band, widely known as ``waterfall''\cite{Ronning2005}. The waterfall-like behavior is found to be rather common in cuprates, and great efforts have been made to understand the nature of this feature\cite{Graf2007,Xie2007,Valla2007,Meevasana2007,Chang2007,Inosov2007,Rienks2014,Byczuk2007,
Tan2007,Zhou2007,Macridin2007,Markiewicz2007,Manousakis2007,Alexandrov2007,Leigh2007,Zhu2008,
Srivastava2008, Zemljic2008,Tan2008,Zhang2008,Moritz2009}. Yet a systematic momentum-dependent study of this high energy anomaly  in the parent compound CCOC is still lacking, and it is still under debate whether it represents intrinsic band structure or not.

In this letter, we present detailed high resolution angle-resolved photoemission measurements on CCOC. Using different experimental geometries and examining carefully the underlying Fermi momentum k$_F$,  we reveal, for the first time, the existence of multiple underlying Fermi surface sheets (labeled as $\alpha$  and $\beta$) in CCOC. The shape of the $\alpha$ underlying Fermi surface seems to be quite similar to the hole-like Fermi surfaces in hole-doped cuprate superconductors, while the $\beta$ underlying Fermi surface is anomalous and seems to form an electron-like pocket around $\Gamma$ point. We also investigate the detailed momentum dependence of the waterfall-like high energy bands along the $\alpha$ underlying Fermi surface. Near the nodal region, the high energy band dispersions resemble an LDA calculated bare band, while in the antinodal region, the high energy bands strikingly exhibit a nearly vertical dispersion. The high energy kink is found to be $\sim$0.9~eV, and does not show much momentum-dependent change within the error bar due to its intrinsic broad feature. Along the $\alpha$  and $\beta$ underlying Fermi surface sheets, we identify two energy scales in the energy distribution curves (EDCs) at underlying k$_F$: the peak maximum, which shows a \emph{d}-wave-like form, and the spectral weight onset position, which is found to be $\sim$0.2~eV and nearly isotropic. The low energy peak dispersion and the corresponding \emph{d}-wave-like gap is consistent with previous ARPES results\cite{Ronning1998}. The isotropic $\sim$0.2~eV gap determined by the spectral weight onset position is consistent with previous STM results\cite{Ye2013}, which indicates the compound is insulating and fully gapped.  Our results provide new information on the electronic structure of the CCOC parent compound, and indicate that the strong electron correlation nature of the Mott insulator may be responsible for the complexity of the underlying Fermi surfaces and the anomalous momentum dependence of the high energy bands.



Figure 1 shows the constant energy contours of Ca$_2$CuO$_2$Cl$_2$ at different binding energies measured under two distinct experimental geometries. To avoid an obvious sample charging, the samples were measured at 86 K. Under both experimental geometries, no spectral weight is observed at the Fermi level (not shown in Fig. 1), consistent with the insulating nature of CCOC. At a binding energy of 0.25 and 0.3 eV, spectral weight begins to emerge around ($\pi/2$,$\pi/2$). At a binding energy of 0.4 eV, four separated strong intensity segments appear near the nodal region in the first Brillouin zone (Fig. 1c), which is quite similar to the notable ``Fermi arcs'' in underdoped cuprates. In the geometry I, the intensity of the segments in the first and third quadrants of the first Brillouin zone are relatively stronger than those in the second and fourth quadrants, due to the photoemission matrix element effect\cite{Damascelli2003}. This can be seen from the fact that, the segments show similar intensity under the experimental geometry II. Each segment grows in area with increasing binding energy and gradually spreads towards the antinodal regions. At a first glance, the segments seem to form a simple hole-like pocket around ($\pi$,$\pi$) point at a binding energy of 1.0 eV and above in Fig. 1a. For a better view of the first Brillouin zone, the original data in Fig. 1a is symmetrized, as shown in Fig. 1c. A close inspection reveals some weak spectral weight along (0,0)-(0,$\pi$) direction at a binding energy of 0.8 eV and above, which is more obvious under geometry II (Fig. 1b), indicating the existence of a second sheet of underlying Fermi surface, which is illustrated in Fig. 2c.

Figure 2 shows the detailed momentum evolution of the band structures in CCOC. The band images in Fig. 2 d and e are measured under the  experimental geometry I along the vertical and horizontal cuts labeled in Fig. 2a, respectively. The band images in Fig. 2 f and g are measured under the geometry II along the vertical and horizontal cuts labeled in Fig. 2b, respectively. For each momentum cut, there are two features to be noticed: one is the low energy charge transfer band feature located approximately at 0.5-0.9 eV, the other is the rapidly dispersing high energy feature above $\sim$0.9 eV. Similar results were also reported before\cite{Ronning2005}. From the nodal region to the antinodal region, the band top gradually shifts downwards, and a fair amount of spectral weight is transferred from the low energy band to the high energy band. For each band, the underlying k$_F$ is extracted by energy distribution curve (EDC) and momentum distribution curve (MDC) analysis methods. For the EDC analysis, the corresponding momentum where the EDC peak position is  closest to the Fermi level, is defined as the EDC underlying k$_F$.  For the MDC analysis, the band top can be inferred from the MDC-fitted band dispersion which starts to become vertical;  the corresponding momentum is defined as the MDC underlying k$_F$. In our case, the two methods give quantitatively similar underlying k$_F$. By tracing the underlying k$_F$ from two independent measurements under two different experimental geometries (Fig. 2a and Fig. 2b), the underlying Fermi surface contours are extracted and shown in Fig. 2c. For the first time, with different experimental geometry to check matrix element, improved energy and angle resolution and detailed mapping in the momentum space, two underlying Fermi surface sheets are clearly revealed, labeled as $\alpha$ and $\beta$ in Fig. 2c. From the band structure evolution, we can identify that the $\alpha$ sheet is  ``hole-like" and centered around ($\pi$,$\pi$), the $\beta$ sheet is ``electron-like" and centered around $\Gamma$, which indicate a balance of the hole number and electron number in this system, as expected in the half-filled parent compound.

Figure 3 shows the MDC-fitted high energy band dispersions along the $\alpha$ underlying Fermi surface sheet. Fig. 3 a-j shows the band structures along the momentum cuts marked in Fig. 3k. The original data in a and j are symmetrized for a better view of the high energy dispersion. Fig. 3l shows the MDC-derived dispersions of the bands in Fig. 3 a-j. From the momentum dependent behavior of the MDC dispersions, two distinct momentum regions can be identified. Near the nodal region, the high energy kink position is found to be $\sim$0.9 eV and does not show much change from cut 1 to 6. The high energy waterfall-like dispersion matches relatively well with the LDA calculated bare band from cut 1 to 3. Near the antinodal region, the high energy kink can not be discerned, and the high energy dispersion shows nearly no sensitivity of energy, dispersing almost vertically, which is quite different from the behavior near the nodal region . From the band top momenta of the MDC dispersions, the underlying k$_F$ of the $\alpha$ sheet underlying Fermi surface are obtained and shown in Fig. 3k. It matches well with the intensity plot of the constant energy contour at 0.3 eV. We note that the bands of the $\beta$ underlying Fermi surface sheet is too weak to be discerned from the bands in a-j  along the selected cuts under the geometry I. It becomes relatively stronger when measured in the geometry II (Fig. 2b).  Fig. 3m shows the extracted MDC width of the bands in a-j. In the intermediate region between nodal and antinodal regions, the high energy spectral peaks are too broad to be fitted well, leading to a relatively large width, like in the case of cut 4 and 5.  The drop at high energy $\sim$0.9 eV in the MDC width can be seen from the cuts 1 to 6, which is consistent with the results in Fig. 3l, indicating the existence of a kink feature. As mentioned above, MDC fitting at high energies does not work well in the intermediate region. Getting rid of cut 4 and 5, we observe that the width of the high energy waterfall bands in different cuts coincides well with each other, despite the large change in the width of low energy bands.

Figure 4 shows the momentum dependence of two energy scales along the underlying Fermi surface in CCOC. The typical EDCs of the band image in Fig. 3a and Fig. 3j are shown in Fig. 4 a and Fig. 4b, respectively. The EDCs closest to the Fermi level are marked in red, which represent the EDCs at underlying k$_F$. In the nodal EDCs (Fig. 4a), a broad peak feature can be observed which disperses fast towards the low energy band top $\sim$0.5 eV, and then disperses back while passing through the underlying k$_F$. In the antinodal EDCs (Fig. 4b), only a broad hump near $\sim$0.8 eV and a step-like feature can be discerned. In Fig. 4d and 4f, the EDCs along the $\alpha$ and $\beta$ underlying Fermi surface sheets (Fig. 4c) are shown. The EDCs are vertically offset for clarity. Fig. 4 e and 4g shows the logarithm plot of the original EDCs in Fig. 4d and 4f, respectively. From Fig. 4d-g, two energy scales can be identified:  the low energy peak position and the spectral weight onset position. The momentum dependence of the two energy scales is shown in Fig. 4h. It can be seen that, the gap size obtained from the peak positions is anisotropic from nodal to antinodal regions and follows an offset \emph{d}-wave-like gap form $\Delta=\Delta_{n} + \Delta_0\cos(2\Phi)$ with $\Delta_{n}$ representing the nodal gap (black line in Fig. 4h)\cite{Ronning1998,PYY2013}. The gap size obtained from the spectral weight onset position is $\sim$0.2 eV and basically isotropic in the momentum space.


To the best of our knowledge, multiple underlying Fermi surface sheets are clearly revealed for the first time in the Mott insulator CCOC. Previously, the steepest descent in the momentum distribution $n(k)$, i.e. $|\nabla n(k)|$, is used to determine the underlying k$_F$. From the experimental point of view, there are a number of issues associated with this method, making it difficult to extract precisely the underlying k$_F$ point\cite{Ronning1998}.  One is the strong momentum-dependence of the matrix elements, as can be seen in our case in the two different experimental geometries. The second is the need to integrate over a large energy window, leading to a distribution $n(k)$ mixed with contribution both from the low energy and high energy bands. Due to the large background intensity at high energy or near antinodal region, the $n(k)$ lineshape may not reflect the low energy dispersing feature anymore. Here, with improved energy and momentum resolution, the implementation of the EDC and MDC analysis methods provides more reliable results. We also note that, the $\beta$ underlying Fermi surface sheet is not the corresponding shadow Fermi surface of the $\alpha$ sheet because the two sheets are clearly not symmetrical with respect to the antiferromagnetic Brillouin zone boundary beyond the experimental uncertainty.


The peculiar momentum dependence of the high energy waterfall-like bands can not be understood by a simple LDA bare band picture nor \emph{t-J} model. From an experimental point of view, considering the nearly vertical dispersing low energy bands in the antinodal region, the high energy band may be  intrinsic to the cuprate parent compound and may play an important role to connect the low energy charge transfer band with the valence band at higher energy which is assumed to be of strong oxygen \emph{2p} orbital character\cite{Moritz2009}. The multiple underlying Fermi surface nature and the anomalous momentum dependence of the high energy bands reflect the strong electron correlation nature in Mott insulator and need further theoretical study.  Although ARPES probes only the occupied states which cannot explore the upper Hubbard band,  our identification of the nearly isotropic full gap upon the lower Hubbard band may indicate a similar behavior of the upper Hubbard band.  In this case, these may finally lead to an isotropic Mott gap in momentum space. Further experimental investigations are needed to obtain the complete momentum dependent information of the Mott-Hubbard gap.


In conclusion, by taking high resolution ARPES measurements, for the first time, we have clearly revealed multiple underlying Fermi surface sheets and the anomalous momentum dependence of the high energy waterfall-like bands in the parent Mott insulator CCOC. We also identify two distinct energy scales in the spectra:  the low energy peak dispersion which shows a \emph{d}-wave-like form and a nearly isotropic lower Hubbard band gap. Our observations provide new information on the electronic structure of the Mott insulator and indicate that the strong electron correlation effect is important to understand the anomalous physical properties of the cuprate parent compound.

\vspace{3mm}

{\bf Methods}

High quality Ca$_2$CuO$_2$Cl$_2$ (CCOC) single crystals are grown by the flux method\cite{Hiroi1994}. Polycrystalline samples are synthesized by mixing the powders of CaO and CuCl$_2$ with a molar ratio of 2:1 in an alumina crucible and heating at 1,073 K  for 24 h with intermediate grindings. The CCOC precursor is then heated to 1,203 K at a ramp rate of 60 K/h and kept there for 10 h. Finally, CCOC single crystals are obtained by cooling  down to room temperature at a ramp rate of 60 K/h.

The ARPES measurements are performed on our lab photoemission system equipped with a Scienta DA30L electron energy analyzer and a Helium discharge lamp with a photon energy 21.218 eV\cite{Liu2008}. The incident light is partially polarized, with a major s polarization. The overall energy resolution was set at 10 meV and the angular resolution is $\sim$0.3$^\circ$. The Fermi level is referenced by measuring the Fermi edge of a clean polycrystalline gold that is electrically connected to the sample. The samples were cleaved \emph{in situ} and measured in vacuum with a base pressure better than 5$\times$10$^{-11}$ Torr.

\vspace{3mm}

$^{*}$Corresponding author: XJZhou@iphy.ac.cn.

\textbf{}

\vspace{3mm}

\noindent {\bf Acknowledgement}\\
XJZ thanks financial support from the National Key Research and Development Program of China (2016YFA0300300), the National Natural Science Foundation of China (11334010 and 11534007), the National Basic Research Program of China (2015CB921000) and the Strategic Priority Research Program (B) of Chinese Academy of Sciences (XDB07020300).

\vspace{3mm}

\noindent{\bf Additional information}\\
Correspondence and requests for materials should be addressed to X.J.Z.

\newpage

\begin{figure*}[tbp]
\begin{center}
\includegraphics[width=1.0\columnwidth,angle=0]{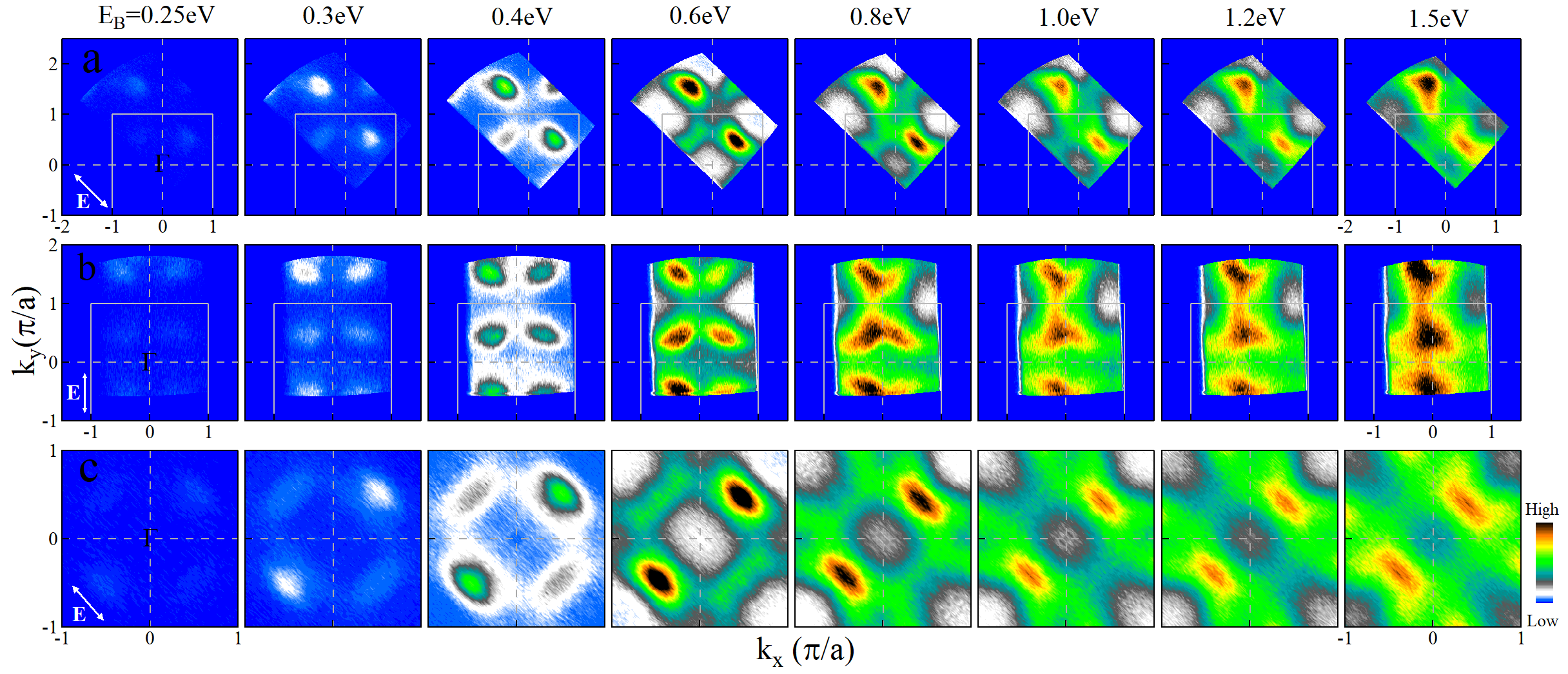}
\end{center}

\caption{{\bf Constant energy contours of Ca$_2$CuO$_2$Cl$_2$ measured at 86~K.} (a-b) Constant energy contours of Ca$_2$CuO$_2$Cl$_2$ measured in geometry I and geometry II, respectively, obtained by integrating the photoemission spectral weight over a 30 meV energy window at different binding energies (E$_B$) of 0.25 eV, 0.3 eV, 0.4 eV, 0.6 eV, 0.8 eV, 1.0 eV, 1.2 eV and 1.5 eV (from left column to right column). In geometry I, the cut orientation is parallel to the (0,0)-($\pi$,$\pi$) direction. In geometry II, the cut orientation is parallel to the (0,0)-($\pi$,0) direction, which is Cu-O-Cu bond direction. The electric field direction of the incident light in each geometry is shown in the leftmost  panels. The first Brillouin zone is indicated by a gray square. (c) Constant energy contours of Ca$_2$CuO$_2$Cl$_2$ in geometry I obtained by a four-fold symmetrization of the original data  from (a) within the region between (0,0)-($\pi$,$\pi$) line and (0,0)-(-$\pi$,$\pi$) line. Only the first Brillouin zone is shown here. }

\end{figure*}

\begin{figure*}[tbp]
\begin{center}
\includegraphics[width=1.0\columnwidth,angle=0]{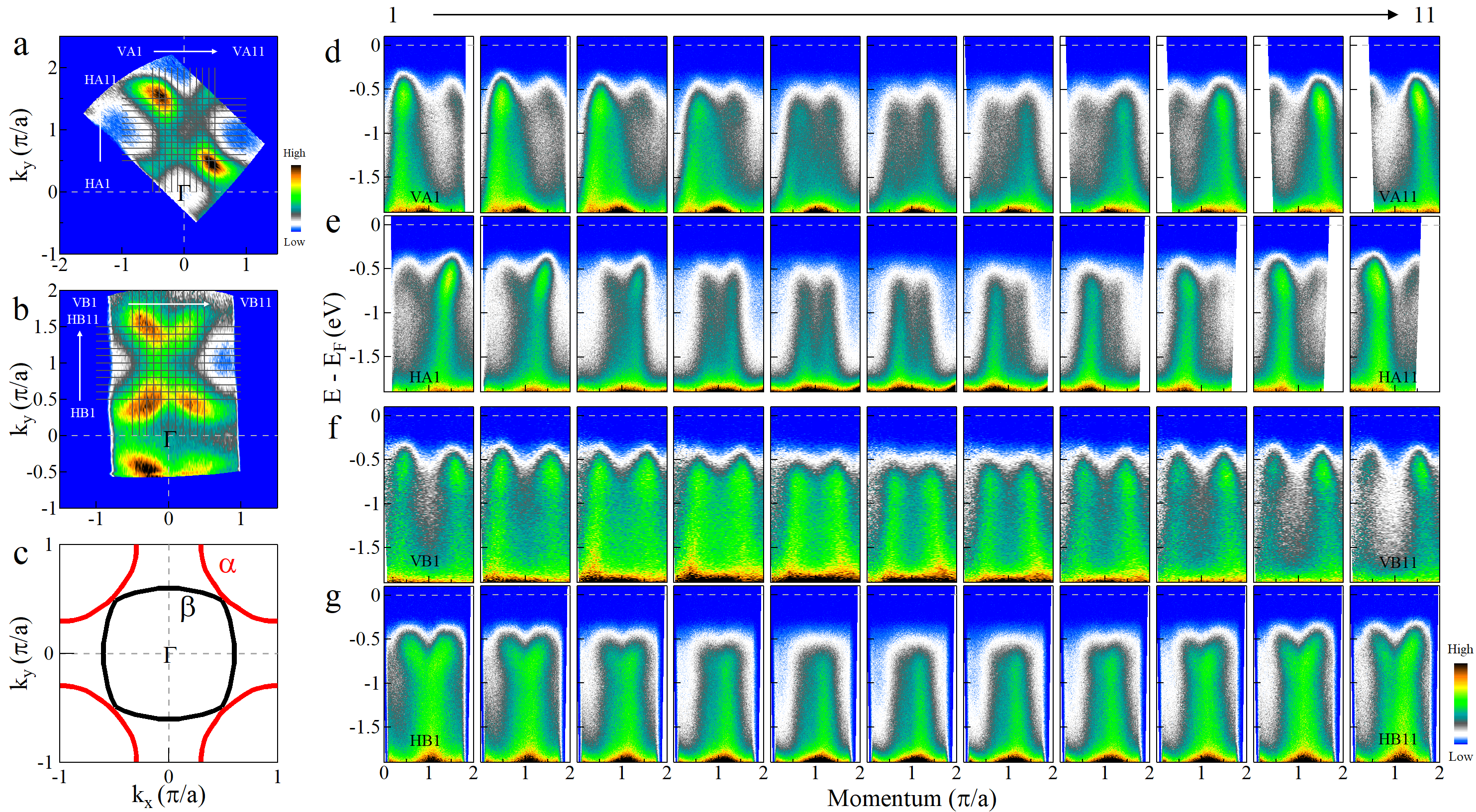}
\end{center}

\caption{{\bf Momentum dependence of the band structure in Ca$_2$CuO$_2$Cl$_2$.} (a) Constant energy contour of Ca$_2$CuO$_2$Cl$_2$ measured in geometry I (same as Fig. 1a) at a binding energy of 0.7 eV, with vertical momentum cuts labeled from VA1 to VA11,  and horizontal momentum cuts labeled from HA1 to HA11. (b) Constant energy contour of Ca$_2$CuO$_2$Cl$_2$ measured in geometry II (same as Fig. 1b) at a binding energy of 0.7 eV, with vertical momentum cuts labeled from VB1 to VB11, and horizontal momentum cuts labeled from HB1 to HB11. (c) Multiple underlying Fermi surface sheets obtained by tracing the underlying ``Fermi momentum k$_F$" which is determined by MDC (momentum distribution curve) analysis and EDC (energy distribution curve) analysis of the bands in (d-g). The MDC fitting and EDC analyses quantitatively give similar results. We note that, near the antinodal region where the EDC analysis does not work well, only MDC fitting results are used. The red curve represents  underlying Fermi surface $\alpha$ while the black curve represents  underlying Fermi surface $\beta$. (d) Band structure along various momentum cuts from VA1 to VA11. (e) Band structure along various momentum cuts from HA1 to HA11. (f) Band structure along various momentum cuts from VB1 to VB11. (g) Band structure along various momentum cuts from HB1 to HB11.
}
\end{figure*}

\begin{figure*}[tbp]
\begin{center}
\includegraphics[width=1.0\columnwidth,angle=0]{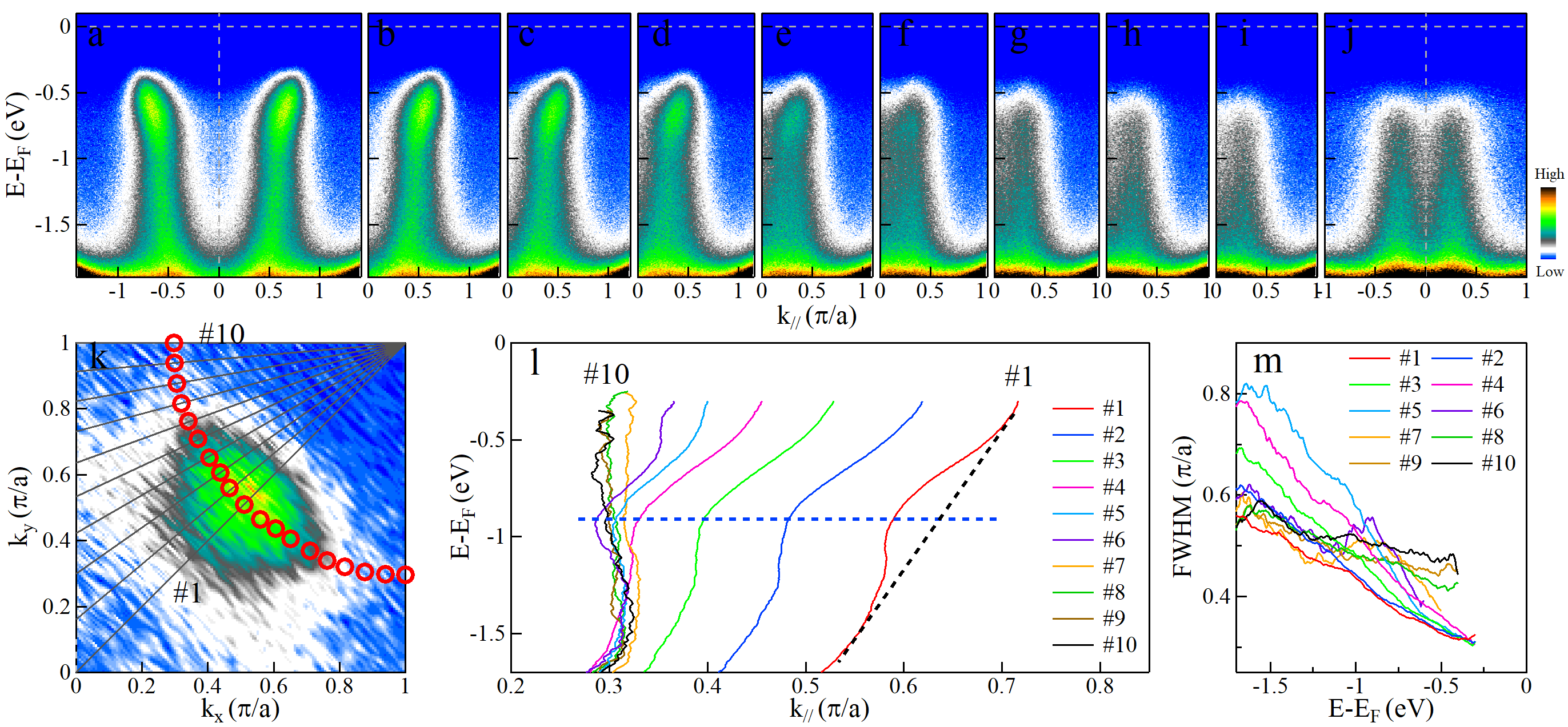}
\end{center}

\caption{{\bf Momentum dependence of the band structure along the $\alpha$ sheet of the underlying Fermi surface and the corresponding MDC-fitted  dispersions.} (a-j) Band structure along momentum cuts from 1 to 10 shown in (k).  The momentum cuts are nearly perpendicular to the underlying Fermi surface sheet $\alpha$. The data in (a) is obtained by symmetrization around $\Gamma$ point. The data from (b) to (i) are original data. The data in (j) is obtained by symmetrization around (0,$\pi$). (k) Constant energy contour of Ca$_2$CuO$_2$Cl$_2$ at a binding energy of 0.3 eV, with momentum cuts labeled from 1 to 10. The red circles represent the underlying k$_F$ determined for each cut. It matches well with the intensity plot of the low energy contours. (l) MDC-derived band dispersions for the momentum cuts 1 to 10.  The blue dashed line is a guideline to the high energy kink position. The black dashed line represents LDA calculated bare band. (m) Extracted MDC width (Full width at half maximum, FWHM).
}
\end{figure*}

\begin{figure*}[tbp]
\begin{center}
\includegraphics[width=1.0\columnwidth,angle=0]{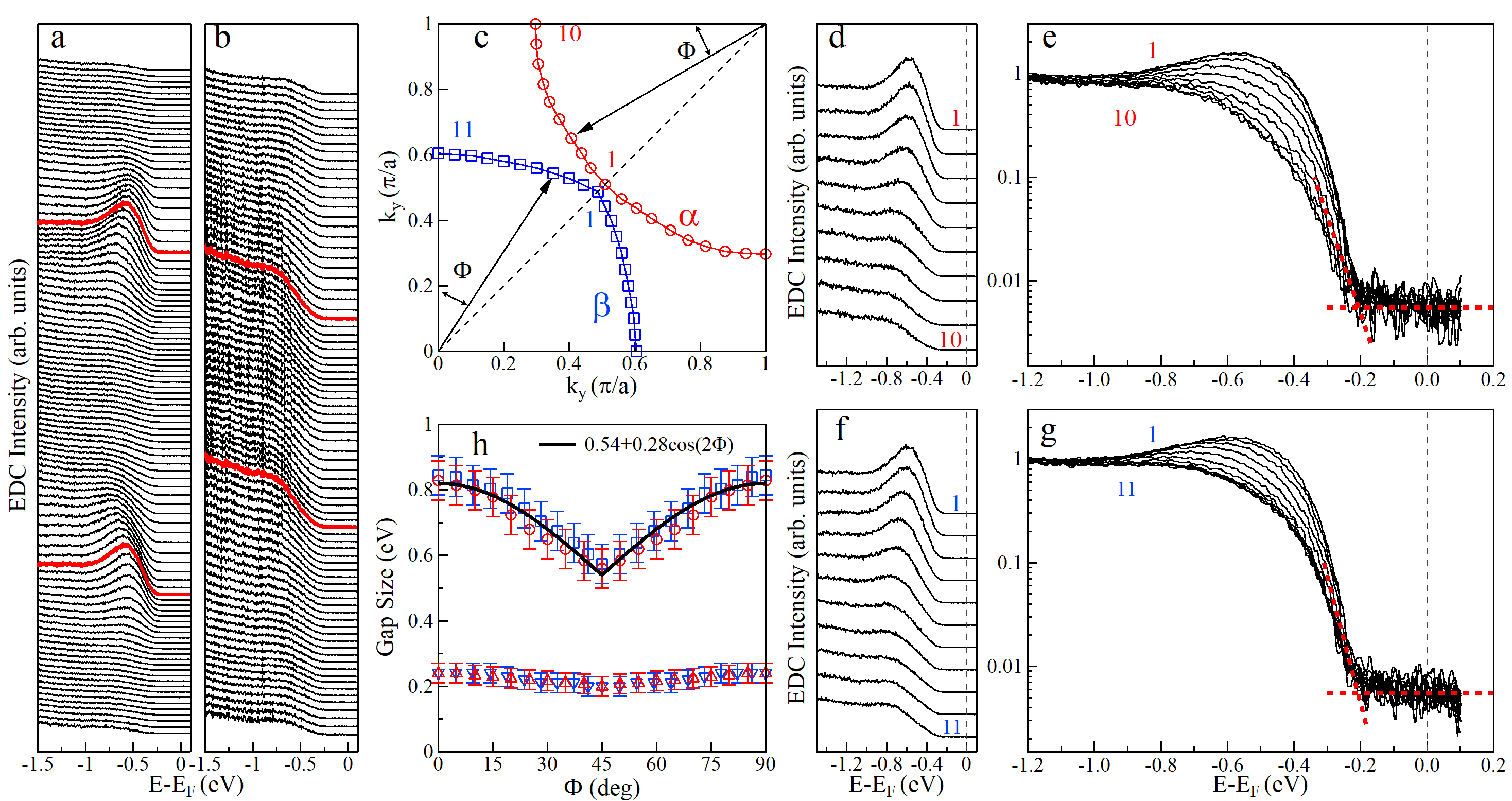}
\end{center}

\caption{{\bf Determination of the two energy scales in Ca$_2$CuO$_2$Cl$_2$.}  (a) Typical EDCs along (-$\pi$,-$\pi$)-(0,0)-($\pi$,$\pi$) direction. The data are symmetrized around $\Gamma$ point (corresponding to Fig. 3a). The red curves represent the EDCs at the underlying k$_F$. (b) Typical EDCs along (-$\pi$,$\pi$)-(0,$\pi$)-($\pi$,$\pi$) direction. The data are symmetrized around (0,$\pi$) (corresponding to Fig. 3j). The red curves represent the EDCs at the underlying k$_F$. (c) The location of the underlying k$_F$ along the underlying Fermi surface sheets $\alpha$ and $\beta$. (d) Corresponding EDCs at underlying Fermi momenta from 1 to 10 along the $\alpha$ underlying Fermi surface in (c). (e) The logarithm plot of the EDCs in (d), showing how the spectral weight onset position is determined. The red dashed lines are guidelines to show the existence of a full gap. (f) Corresponding EDCs at underlying Fermi momenta from 1 to 11 along the $\beta$ underlying Fermi surface in (c). (g) The logarithm plot of the EDCs in (f). (h) Extracted two kinds of energy scales as a function of momentum angle ($\Phi$) that is defined in (c). Note that for the $\alpha$ underlying Fermi surface, the $\Phi$ is defined from ($\pi$,$\pi$) point, while for the $\beta$ underlying Fermi surface,  $\Phi$ is defined from $\Gamma$ point. The red circle and triangle represent the peak position and spectral weight onset position along the $\alpha$ underlying Fermi surface, respectively. The blue square and inverted triangle represent the peak position and spectral weight onset position along the $\beta$ underlying Fermi surface, respectively. The black line depicts the offset \emph{d}-wave form $\Delta=\Delta_{n} + \Delta_0\cos(2\Phi)$ with $\Delta_n=0.54$ eV and $\Delta_{0}=0.28$ eV, in which $\Delta_{n}$ represents the nodal peak position.
}

\end{figure*}

\end{document}